# High-performance solid-state electrochemical thermal switches with earth-abundant cerium oxide

Ahrong Jeong[1#], Mitsuki Yoshimura[2#], Hyeonjun Kong[2], Zhiping Bian[2], Jason Tam[3], Bin Feng[3], Yuichi Ikuhara[3], Takashi Endo[1], Yasutaka Matsuo[1], and Hiromichi Ohta[1,*]

[1] *Research Institute for Electronic Science, Hokkaido University, N20W10, Kita, Sapporo 001-0020, Japan*
[2] *Graduate School of Information Science and Technology, Hokkaido University, N14W9, Kita, 060-0814 Sapporo, Japan*
[3] *Institute of Engineering Innovation, The University of Tokyo, 2-11-16 Yayoi, Bunkyo, Tokyo, 113-8656 Japan*

[#]Contributed equally
*Corresponding author: Hiromichi Ohta (hiromichi.ohta@es.hokudai.ac.jp)

**Abstract**

Thermal switches, which electrically turn heat flow on and off, have attracted attention as thermal management devices. Electrochemical reduction/oxidation switches the thermal conductivity ($\kappa$) of active metal oxide films. The performance of the previously proposed electrochemical thermal switches is low; on/off $\kappa$-ratio is mostly less than 5 and $\kappa$-switching width is less than 5 W/mK. We used $CeO_2$ thin film as the active layer deposited on a solid electrolyte YSZ substrate. When the $CeO_2$ thin film was reduced once (off-state) and then oxidized (on-state), $\kappa$ was about 2.2 W/mK in the most reduced state, and $\kappa$ increased with oxidation to 12.5 W/mK (on-state). This reduction (off-state)/oxidation (on-state) cycle was repeated 100 times and the average value of $\kappa$ was 2.2 W/mK after reduction (off-state) and 12.5 W/mK after oxidation (on-state). The on/off $\kappa$-ratio was 5.8 and $\kappa$-switching width was 10.3 W/mK. The $CeO_2$-based solid-state electrochemical thermal switches would be potential devices for thermal shutters and thermal displays.

**Introduction**

Due to low energy conversion efficiency, two-thirds of primary energy such as oil and coal is lost as waste heat. In particular, low- to medium-temperature (100−300 °C) waste heat is the most difficult to reuse.(*1*) Thermoelectric energy conversion technology, which directly converts a temperature difference into electricity, is one solution, but its efficiency is low in this temperature range in air.(*2, 3*) On the other hand, thermal management technologies such as thermal diodes and thermal transistors (hereinafter referred to as thermal switches) have recently attracted attention.(*4*) Thermal diodes rectify heat flow and thermal switches turn heat flow on and off. We expect that thermal displays that visualize heat contrast using infrared cameras can be realized using thermal switches. Similar to liquid crystal displays, the transmission of backlight (infrared energy from waste heat) can be controlled by thermal switches. Therefore, thermal switches may be useful for reusing waste heat as an infrared source.

Several types of thermal transistors have been proposed. First, theorists proposed thermal switches that use heat-induced modulation of thermal conductivity ($\kappa$) modulation.(*5-7*) In 2014, Ben-Abdallah and Biehs proposed electrical switching of the $\kappa$ of $VO_2$, which exhibits an insulator-to-metal transition at 68 °C in bulk form.(*6*) Unfortunately, the idea was not realized because the $\kappa$ of $VO_2$ does not change when the temperature reaches the transition temperature.(*8, 9*) Since then, several thermal transistors have been proposed that exploit the $\kappa$ modulation by the electrochemical reaction of active materials: $LiCoO_2/Li_{1-\delta}CoO_2$ (*10*), $MoS_2/Li_xMoS_2$ (*11*), $SrCoO_3/SrCoO_{2.5}/HSrCoO_{2.5}$ (*12*), $MoS_2/MoS_2$-organic molecular hybrid (*13*), $SrCoO_3/SrCoO_2$ (*14-16*), $La_{0.5}Sr_{0.5}CoO_3/La_{0.5}Sr_{0.5}CoO_{2.5}$ (*17*), and $LaNiO_3/LaNiO_{3-\delta}$ (*18*). In addition, Tomko *et al.*(*19*) reported thermal conductivity switching in topologically networked bio-inspired materials. Castelli *et al.*(*20*) reported three-terminal magnetic thermal switches that control heat flow by changing the touch/removal of the metal bridge using a magnet. Liu *et al.*(*21*) demonstrated low-voltage controlled thermal switching in antiferroelectric $PbZrO_3$ thin films. Li *et al.*(*22*) developed a three-terminal thermal switch that electrostatically controls the $\kappa$ of an organic molecule-based self-assembled monolayer. In addition, Hartquist *et al.*(*23*) demonstrated that the deformation of polymer materials can be used to tune thermal conductivity. Among these various proposed thermal switches, in this study, we focused on oxide-based solid-state electrochemical thermal switches.(*14*)

Oxide-based solid-state electrochemical thermal switches have several disadvantages,

including low operating speed, high operating temperature, low performance, and the presence of rare metal elements. First, compared to electrostatic modulation,(*22*) electrochemical modulation takes time to turn thermal switches on and off. Although it is necessary to modulate the thermal conductivity of thick active materials for practical applications such as thermal displays, electrostatic modulation can only modulate the interface between the gate and the active materials. With electrochemical thermal switches, the thermal conductivity of the entire film can be modulated. This is one advantage of electrochemical thermal switches. Second, the operating temperature of solid-state electrochemical thermal switches is high (~280 °C) compared to liquid electrolyte-based electrochemical thermal switches, which can be operated at room temperature. However, we want to use the low-to-medium temperature (100−300 °C) waste heat exhausted into the air as the backlight of thermal displays, so a relatively high operating temperature is not a problem. In addition, the thermal switches must be stable over the temperature range and must not contain any liquid. Since most oxide materials are generally stable in the temperature range in air, we chose oxides as the active materials. Third, the performance such as the on/off $\kappa$-ratio and $\kappa$-switching width of oxide-based solid-state electrochemical thermal switches is lower compared to the other types of thermal switches. According to our previous studies, materials with high electrical conductivity are promising as active materials to improve the $\kappa$-switching width of solid-state thermal switches, because such materials have high electron thermal conductivity.(*15*) Recently, we have focused on $LaNiO_3$ as an active material of solid-state thermal switches because $LaNiO_3$ exhibits very high electrical conductivity (~10,000 S cm$^{-1}$) and high $\kappa$ (11 W m$^{-1}$ K$^{-1}$ (*24*)) at room temperature. We have achieved $LaNiO_3$-based thermal switches with a large $\kappa$-switching width (~4.3 W m$^{-1}$ K$^{-1}$) by exploiting its high electron thermal conductivity of ~3.1 W m$^{-1}$ K$^{-1}$.(*18*) Although this widens the $\kappa$-switching width, the width is still less than 5 W m$^{-1}$ K$^{-1}$. Therefore, we are still searching for oxide-based solid-state electrochemical thermal switches with a large on/off $\kappa$-ratio and a much wider $\kappa$-switching width. Fourth, the previously reported oxide-based thermal switches contain rare metal elements of Co and Ni. The use of rare metal elements is not suitable for sustainable development.

In this study, we focused on the electrochemical reduction/oxidation of $CeO_2$ as an active material for thermal switches. $CeO_2$ is an abundant material that is widely used in practical applications such as polishing powders, catalysts, and sunscreens. The crystal structure of $CeO_2$ is a simple cubic fluorite (space group: $Fm\bar{3}m$), with a lattice parameter of 0.541 nm. $CeO_2$ films can be grown heteroepitaxially on YSZ single crystals ($a$ = 0.515 nm) with a lattice mismatch of about +5%. Bulk $CeO_2$ exhibits a high $\kappa$ value (~14 W m$^{-1}$ K$^{-1}$) at room temperature,(25-27) which is higher than that of $SrCoO_3$ (3.8 W m$^{-1}$ K$^{-1}$ (28)) and $LaNiO_3$ (11 W m$^{-1}$ K$^{-1}$ (24)). Moreover, $CeO_2$ can be reduced thermochemically (29-33) and electrochemically.(34) Oxygen-deficient homologous phases of $Ce_nO_{2n-2}$ ($n \geq 4$, integer) are known.(29-33) In 2014, Khafizov *et al*. obtained theoretical results indicating that $CeO_{1.735}$ ($n$ ~7 in $Ce_nO_{2n-2}$) would have a very low $\kappa$ value (~1.2 W m$^{-1}$ K$^{-1}$),(35) which is close to the amorphous limit for thermal transport ($\kappa$ ~0.9 W m$^{-1}$ K$^{-1}$).(36) Therefore, we expected that $CeO_2$-based thermal switches would have a large on/off $\kappa$-ratio > 10 and a large $\kappa$-switching width ~13 W m$^{-1}$ K$^{-1}$, which would be outstanding among the reported oxide-based thermal switches.

Here, we show that $CeO_2$-based solid-state electrochemical thermal switches exhibit a relatively large on/off $\kappa$-ratio of 5.8 and a wide $\kappa$-switching width of ~10.3 W m$^{-1}$ K$^{-1}$ (on-state: 12.5 W m$^{-1}$ K$^{-1}$, off-state: 2.2 W m$^{-1}$ K$^{-1}$). This performance is outstanding among reported oxide-based solid-state electrochemical thermal switches. Since $CeO_2$ is an earth-abundant material, the present $CeO_2$-based solid-state electrochemical thermal switches would have a potential for use in practical thermal management devices such as thermal displays.

**Results**
**Electrochemical reduction/oxidation of cerium oxide films**
First, we fabricated $CeO_2$-based thermal switches (**Fig. 1**). Details of the device fabrication procedure are described in **Supplementary Materials S1** and the **Methods** section. **Figure 1A** shows the device structure of a $CeO_2$-based solid-state electrochemical thermal switch, which consists of a 103-nm-thick $CeO_2$ epitaxial film on a 0.5-mm-thick YSZ substrate sandwiched between Pt films. The electrochemical reduction/oxidation treatments were performed at 280 °C in the air (**Fig. 1B**). A 5 mm × 5 mm device was placed on a heater (280 °C) and a negative/positive voltage was applied to the top Pt electrode. The trilayer structure was visible in the cross-sectional transmission electron microscopy (TEM) image (**Fig. 1C**). The selected area electron diffraction pattern (**Fig. 1D**) indicates that the $CeO_2$ film was heteroepitaxially grown on YSZ with a cube-on-cube relationship.

The electrochemical reduction/oxidation treatments were performed by applying a constant current of −10 μA/+10 μA (**Figs. 2A** and **2D**). These treatments were performed by applying an electron density $Q$ of $1 \times 10^{21}$ cm$^{-3}$ each time. During electrochemical reduction (**Fig. 2A**), the absolute value of the applied voltage remains almost constant (~3.3 V). We repeated the reduction treatment until the total $Q$ reached $-5.5 \times 10^{22}$ cm$^{-3}$, corresponding to a total reduction time of 2255 s. In contrast, the voltage required for initial oxidation was negative, reflecting spontaneous oxidation (**Fig. 2D**). When $Q$ exceeds ~$7 \times 10^{21}$ cm$^{-3}$, the required voltage increases dramatically from ~0 V to ~5 V. The required voltage became saturated when $Q = 1 \times 10^{22}$ cm$^{-3}$ was applied. This corresponds to a total oxidation time of 410 s.

Each time after applying $Q = -1 \times 10^{21}$ cm$^{-3}$/$+1 \times 10^{21}$ cm$^{-3}$, the out-of-plane X-ray diffraction (XRD) pattern of the device was recorded as shown in **Figs. 2B** and **2E**. Upon reduction (**Fig. 2B**), several diffraction peaks of reduced CeO$_{2-\delta}$, labeled **b**, **c**, and **d** appear sequentially in addition to the diffraction peak of CeO$_2$ (labeled **a**). These phases are heteroepitaxially grown on the YSZ substrate without any strain (**Supplementary Material S2**). We then analyzed the homologous phases of Ce$_n$O$_{2n-2}$ ($n \geq 4$, integer) as a function of $\delta$ in CeO$_{2-\delta}$. Using the reported lattice parameters (*32*) of CeO$_2$, CeO$_{1.832}$, and CeO$_{1.735}$, we found a linear relationship between the $\delta$ and the lattice parameter (**Supplementary Material S3**). The linear relationship was then extended to $\delta = 0.5$. The linear relationship, we clarified revealed that the **a**, **b**, **c**, and **d** phases are CeO$_2$ ($n = \infty$), Ce$_9$O$_{16}$, Ce$_3$O$_5$, and Ce$_2$O$_3$, respectively. Upon oxidation (**Fig. 2E**), the diffraction peak of reduced CeO$_{2-\delta}$ (**c** phase) shifts to a larger $q_z/2\pi$ value and eventually disappears. We further clarified that the reduction of Ce$^{4+}$ to Ce$^{3+}$ occurs after the electrochemical reduction treatment (**Supplementary Material S4**).

If the electrochemical reduction of CeO$_2$ obeys Faraday's laws of electrolysis, the following electrochemical reaction occurs:

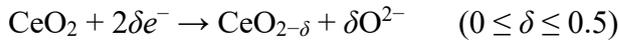

$$\text{CeO}_2 + 2\delta e^- \rightarrow \text{CeO}_{2-\delta} + \delta \text{O}^{2-} \quad (0 \leq \delta \leq 0.5)$$

This indicates that the CeO$_2$ becomes Ce$_2$O$_3$ when the total $Q$ reaches $-2.6 \times 10^{22}$ cm$^{-3}$. However, the XRD results show that there are several phases, CeO$_2$ (**a**) and CeO$_{2-\delta}$ (**b**, **c**, and **d**), probably because the formation energy of homologous phases of Ce$_n$O$_{2n-2}$ ($n \geq 4$, integer) is close to each other and spontaneous oxidation occurs at 280 °C in the air.

**Thermal stability of the reduced cerium oxide film**

When we place the reduced sample on the 280 °C heater without applying a negative current, the film oxidizes within ten minutes. At lower temperatures (100 °C), such spontaneous oxidation does not occur. **Figure 3A** summarizes the change in the out-of-plane XRD patterns of the reduced CeO$_2$-based thermal switch while held at 150 °C in the air. After the reduction ($Q = -1.5 \times 10^{22}$ cm$^{-3}$, 0 h), the diffraction peak of 002 CeO$_{2-\delta}$ ($\delta$ ~0.24, **b**) is seen together with 002 CeO$_2$ (**a**). The peak position of **b** shifts to higher $q_z/2\pi$ side and the peak **b** becomes weaker with time. On the other hand, peak **a** becomes strong with time. After 190 h, peak **b** disappears. The lattice parameter $c$ of the **b** phase gradually decreases with time, while the **a** phase is almost constant (**Fig. 3B**). We calculated the XRD peak intensity ratio ($I$**b**/($I$**a** + $I$**b**)), which reflects the volume fraction of the **b** phase (**Fig. 3C**). At 150 °C, the intensity ratio is ~0.8 at the beginning whereas it decreases almost linearly with time, reaching zero after 190 h. In contrast, the intensity ratio is almost constant at 100 °C. Thus, spontaneous oxidation indeed occurs at 280 °C during electrochemical reduction and this is one of the origins of the coexistence of **a**, **b**, **c**, and **d** phases during reduction.

**Thermal conductivity of oxidized/reduced cerium oxide films**
Next, we measured the thermal conductivity of the CeO$_2$-based thermal switch by a time-domain thermoreflectance (TDTR) method at room temperature. Details of the TDTR measurements are described in **Supplementary Material S5** and the **Methods** section. The typical thermal conductivity of the as-grown CeO$_2$-based thermal switch was 13.5 W m$^{-1}$ K$^{-1}$ (**Fig. S8**), which was similar to that of bulk CeO$_2$ (*25-27*), reflecting the high crystal quality of the CeO$_2$ film. The decay of the TDTR phase signal of the oxidized sample is faster than that of the reduced sample (**Fig. S9**). We simulated the TDTR decay curves and determined the thermal conductivities of the CeO$_2$ films during reduction/oxidation (**Fig. 4A**). During the reduction, the $\kappa$ decreases dramatically with $Q$ and reaches ~2.5 W m$^{-1}$ K$^{-1}$ when $Q \sim -10 \times 10^{21}$ cm$^{-3}$. Thereafter, $\kappa$ is nearly constant (~2.5 W m$^{-1}$ K$^{-1}$). This is larger than Khafizov's calculation (1.2 W m$^{-1}$ K$^{-1}$ (*35*)). On the other hand, when oxidized, the $\kappa$ increases dramatically and reaches ~12.5 W m$^{-1}$ K$^{-1}$ when $Q \sim 11 \times 10^{21}$ cm$^{-3}$. To clarify the origin of the $\kappa$ modulation, we calculated the XRD peak intensity ratio of **a**, **b**, **c**, and **d** phases, which reflects the volume fraction of each phase (**Fig. 4B**). During the reduction, the peak intensity ratio of **a** phase decreases drastically with $Q$, reaching ~0.1 when $Q \sim -10 \times 10^{21}$ cm$^{-3}$. On the other hand, the **b** phase increases and peaks around $Q \sim -10 \times 10^{21}$ cm$^{-3}$ and then decreases. Then, the **c** phase increases and has a peak around $Q \sim -18 \times 10^{21}$ cm$^{-3}$ and then decreases. Finally, the **d** phase increases and saturates around $Q \sim 35 \times 10^{21}$ cm$^{-3}$. During oxidation, the intensity of the **b** phase decreases dramatically and the intensity of the **a** phase increases dramatically. These results indicate that the thermal conductivity correlates only with the volume fraction of the **a** phase.

**Cyclability of the thermal switches**

Next, the cyclability of the thermal switches was tested. The electrochemical reduction/oxidation treatments were performed by applying a constant current of −10 µA/+10 µA. These treatments were performed by applying an electron density $Q$ of −1.5 × $10^{22}$ $cm^{−3}$ for reduction and $Q$ of 0.9 × $10^{22}$ $cm^{−3}$ for oxidation. Treatments were repeated for 100 cycles. We measured changes in the out-of-plane XRD patterns of the $CeO_2$ film during reduction (off-state) and oxidation (on-state) (**Fig. 5A**). In the on-state, only the **a**-phase ($CeO_2$) diffraction peak is observed, whereas in the off-state, diffraction peaks of reduced **b**, **c**, and **d** phases with weak **a** phase are observed in all cases. It is remarkable that we did not observe severe degradation of the crystal structure until 100 times cycling. The TDTR decay curves of the $CeO_2$-based thermal switches were measured every 10 cycles at room temperature (**Fig. S10**), and the thermal conductivity was calculated (**Fig. 5B**). The average thermal conductivity was 2.2 ±0.36 W $m^{−1}$ $K^{−1}$ for the reduced off-state and 12.5 ±0.85 W $m^{−1}$ $K^{−1}$ for the oxidized on-state. The on/off $\kappa$-ratio of the $CeO_2$ layer was 5.8 ±0.85 and the $\kappa$-switching width was 10.3 ±0.98 W $m^{−1}$ $K^{−1}$. The switching of crystallographic phases and thermal conductivity remained stable for at least 100 cycles. This excellent cyclability of the thermal switch would be due to the crystallographic similarity of $CeO_2$ and $CeO_{2−\delta}$ homologous phases.

**Thickness dependence of the switching properties**

To clarify the effect of $CeO_2$ thickness on the thermal switch performance, we fabricated several thermal switches with different $CeO_2$ film thicknesses. Electrochemical reduction of the thermal switches was performed by applying a constant DC current of −10 µA until the total $Q$ reached −1.5 × $10^{22}$ $cm^{−3}$. On the other hand, the oxidation treatment was performed by applying a constant DC current of +10 µA until the total $Q$ reached 0.9 × $10^{22}$ $cm^{−3}$. **Figure S11** summarizes the out-of-plane XRD patterns of the resulting thermal switches. The diffraction peak of **a** phase ($CeO_2$) of the as-grown samples shifts to a higher $q_z/2\pi$ side with increasing thickness. After the reduction treatment, **a**, **b**, and **c** phases are randomly distributed most likely due to the contribution of spontaneous oxidation. After the oxidation treatment, only **a** phase diffraction peaks were observed and the peaks slightly shifted to a higher $q_z/2\pi$ side with increasing thickness. Using the XRD patterns, we extracted the lattice parameter of the as-grown and oxidized $CeO_2$ films (**Fig. 6A**). In the case of the as-grown films, the lattice parameter was larger than that of the bulk (0.541 nm) and decreased with thickness when the thickness is thin (~100 nm). This could be due to the oxygen deficiency of the thinner films. **Figure 6B** summarizes the variation of the thermal conductivity of the $CeO_2$-based thermal switch measured at room temperature with the $CeO_2$ thickness. The thermal conductivity increases dramatically with thickness in both the as-grown and on-state when the thickness is thin, most likely due to an increase in oxygen concentration. Thereafter, the thermal conductivity gradually increases and approaches to the bulk value (14 W m$^{-1}$ K$^{-1}$). From these results, the use of thick film or less oxygen-deficient film is appropriate for realizing solid-state electrochemical thermal switches with large $\kappa$-switching width.

**Discussion**

We have realized high-performance solid-state thermal switches with a relatively high on/off $\kappa$-ratio of 5.8 and a large $\kappa$-switching width of 10.3 W m$^{-1}$ K$^{-1}$ using abundant cerium oxide as the active material. This is an excellent performance compared to other values reported for oxide-based thermal switches (**Fig. 7**). In the oxidized on-state, the $\kappa$ was about 12.5 W m$^{-1}$ K$^{-1}$, and $\kappa$ decreased dramatically to 2.2 W m$^{-1}$ K$^{-1}$ by electrochemical reduction. The oxidized state contains only one $CeO_2$ phase, while the reduced state contains several $CeO_{2-\delta}$ phases ($Ce_nO_{2n-2}$, $n$ = 9, 6, and 4). The switching of crystallographic phases and thermal conductivity were stable for at least 100 cycles. The present $CeO_2$-based solid-state electrochemical thermal switches would have the potential for use in thermal management devices such as thermal displays.

**Materials and Methods**

**Fabrication of the $CeO_2$-based Thermal Switches.** $CeO_2$ thin films were heteroepitaxially grown on (001) YSZ single-crystalline substrates (10 mm × 10 mm, 0.5 mm thickness) by pulsed laser deposition (PLD, KrF excimer laser, ~1.5 J cm pulse, 10 Hz). The substrate temperature was maintained at 800 °C during the film growth. The oxygen pressure was $3 \times 10^{-3}$ Pa. The deposition rate was approximately 10 nm min$^{-1}$. Details of the $CeO_2$ film growth are described elsewhere.(*37, 38*) After the $CeO_2$ film growth, Pt films were sputtered on the $CeO_2$ film surface (80 nm) and the backside (30 nm) of the YSZ substrate. Finally, the substrate was cut into four pieces (5 mm × 5 mm).

**Characterization of the Crystal Structure.** The surface morphology of the as-grown $CeO_2$ films was evaluated using reflection high-energy electron diffraction (RHEED) patterns and topographic atomic force microscopy (AFM) observations. We used high-resolution X-ray diffraction (Cu Kα$_1$ radiation, $\lambda$ = 0.154059 nm, ATX-G, Rigaku Co.) to measure X-ray reflectivity, out-of-plane Bragg diffraction patterns, out-of-plane X-ray rocking curves, and reciprocal space mapping of the CeO$_2$-based thermal switches. To further clarify the crystal structure changes, the atomic arrangements were observed by high-angle annular dark-field (HAADF) scanning transmission electron microscopy (TEM/STEM, JEM-ARM200F, JEOL Co.) operating at 200 kV.

**Redox Treatment.** A CeO$_2$-based thermal switch (5 mm × 5 mm) was placed on a Pt-coated glass substrate and heated at 280 °C in air. Then, we applied a constant current of −10 μA/+10 μA for reduction/oxidation. After applying the current, the sample was immediately cooled to room temperature.

**Measurements of the Thermal Conductivity ($\kappa$).** The $\kappa$ value of CeO$_x$ films perpendicular to the substrate surface was measured by TDTR (PicoTR, NETZSCH Japan). The top Pt film was used as the transducer. The decay curves of the TDTR signals were simulated to obtain $\kappa$. The specific heat capacities of the layers used for the TDTR simulation were Pt: 133 J kg$^{-1}$ K$^{-1}$; CeO$_2$: 358 J kg$^{-1}$ K$^{-1}$; and YSZ: 460 J kg$^{-1}$ K$^{-1}$ (**Table S1**). Details of our TDTR method are described elsewhere.(*28*) Regarding the treatment of the thermal conductivity values, since there were several uncertainties such as the position of the baseline, position of the time zero, and the noise of the signal, error bars of ±15% of the obtained values were used.

**Acknowledgments**

The authors thank Prof. David Cahill for his kind checkup our thermal conductivity data, Ms. Naomi Hirai and Ms. Yuko Mori for their efforts in TEM/STEM analyses and Dr. Yusaku Magari for helpful discussions.

**Funding:**

Japan Society for the Promotion of Science Grants-in-Aid for Scientific Research A 22H00253 (HO), Innovative Areas 19H05791 (HO), and Innovative Areas 19H05788 (BF)

Hokkaido University DX Doctoral Fellowship JPMJSP2119 (ZB)

Advanced Research Infrastructure for Materials and Nanotechnology, Japan and the Ministry of Education, Culture, Sports, Science, and Technology (MEXT) JPMXP1223HK0082 (Hokkaido Univ.)

Crossover Alliance to Create the Future with People, Intelligence and Materials Network Joint Research Center for Materials and Devices

**Author contributions:**

Conceptualization: HO
Methodology: HO
Investigation: HO, AJ, MY, HK, ZB
Visualization: JT, BF, YI, TE, YM
Supervision: HO
Writing—original draft: HO, AJ
Writing—review & editing: HO, AJ

**Competing interests:** The authors declare no competing interests.

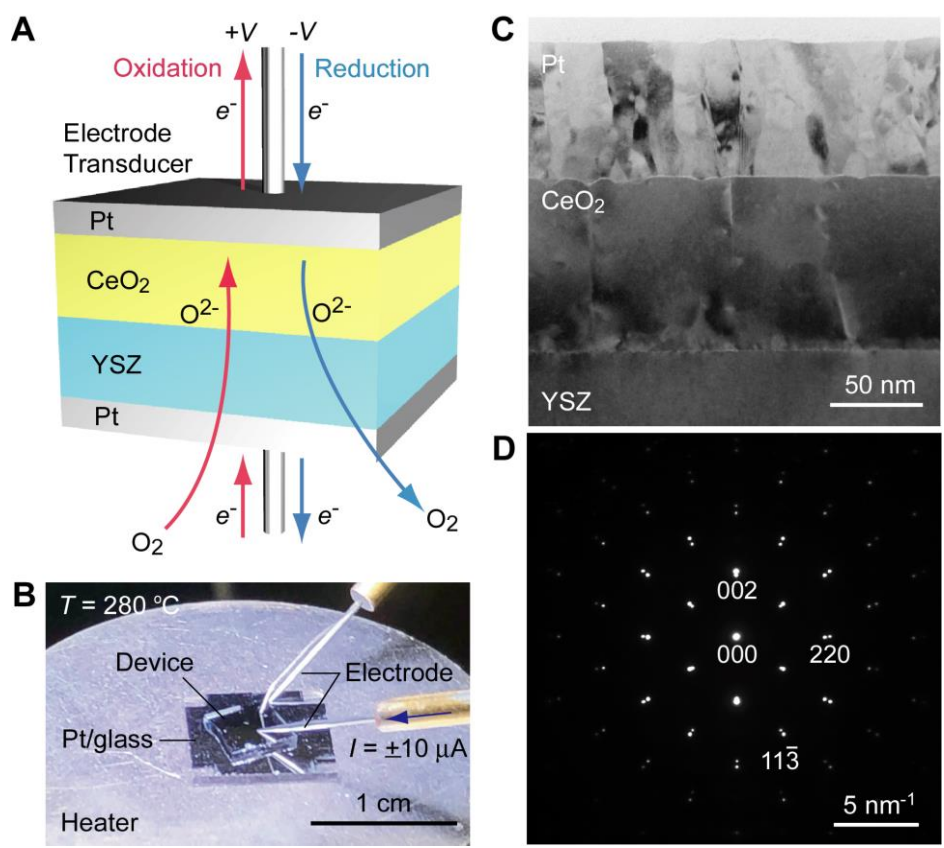

**Figure 1. A CeO$_2$-based solid-state thermal switch.** (**A**) Schematic device structure of a CeO$_2$-based solid-state electrochemical thermal switch, which consists of a 103-nm-thick CeO$_2$ epitaxial film grown on a 0.5-mm-thick YSZ substrate, sandwiched between two Pt films. When a negative/positive voltage is applied to the top Pt electrode, electrochemical reduction/oxidation of the CeO$_2$ layer occurs. The top Pt electrode is also used as a transducer for the thermal conductivity measurements. (**B**) Photograph of a CeO$_2$-based thermal switch operating at 280 °C in the air. The surface area of the device is ~25 mm$^2$. A constant current of +10 μA (−10 μA) was applied for oxidation (reduction). (**C**) Cross-sectional TEM image of the thermal switch. The Pt/CeO$_2$/YSZ trilayer structure is clearly visible. (**D**) Selected area electron diffraction pattern. Diffraction spots of the CeO$_2$ film are seen together with those of the YSZ. The crystallographic orientation is (001)[110]$_{CeO2}$ ∥ (001)[110]$_{YSZ}$.

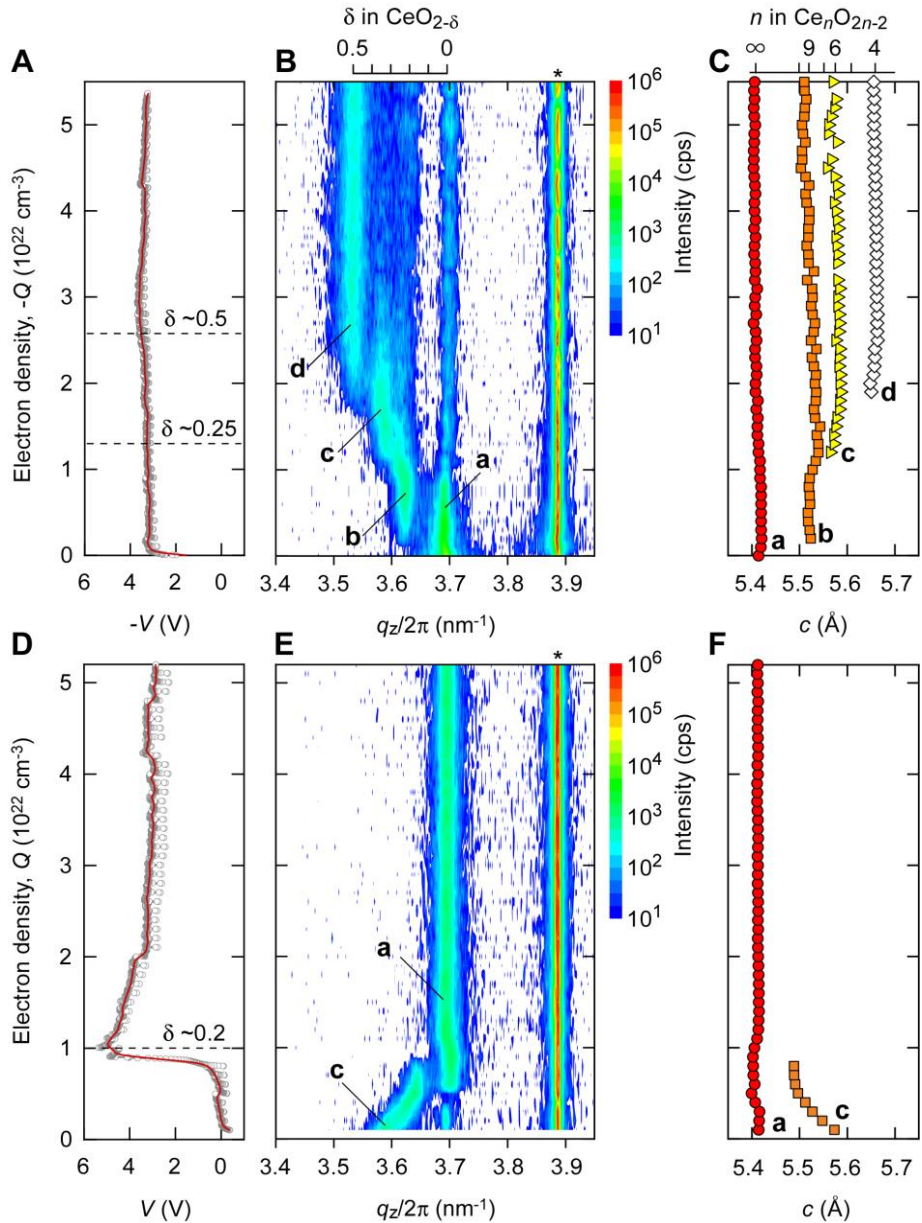

**Figure 2. Electrochemical reduction/oxidation treatment of a $CeO_2$-based solid-state electrochemical thermal switch.** (**A**, **D**) Changes in the applied voltage during reduction and oxidation treatments. During electrochemical reduction, the absolute value of the applied voltage is almost constant (~3.3 V). In contrast, the applied voltage increases dramatically from 0 V to 5 V during oxidation. (**B**, **E**) Changes in the out-of-plane XRD patterns. After reduction treatment, the diffraction peak of 002 $CeO_{2-\delta}$ (labeled **b**, **c**, and **d**) appears sequentially in addition to the diffraction peak of 002 $CeO_2$ (labeled **a**) with 002 YSZ (*). After oxidation treatment, the diffraction peak of 002 $CeO_{2-\delta}$ (**c**) shifts to the larger $q_z/2\pi$ side and finally disappears. (**C**, **F**) Lattice parameters, $c$, of the $CeO_2$ (**a**) and $CeO_{2-\delta}$ (**b**, **c**, and **d**) extracted from the out-of-plane XRD patterns. The $c$ of the $CeO_2$ (**a**) is almost constant (0.541 nm). From the $c$, **a**, **b**, **c**, and **d** phases are found to be $CeO_2$, $Ce_9O_{16}$, $Ce_6O_{10}$, and $Ce_2O_3$, respectively.

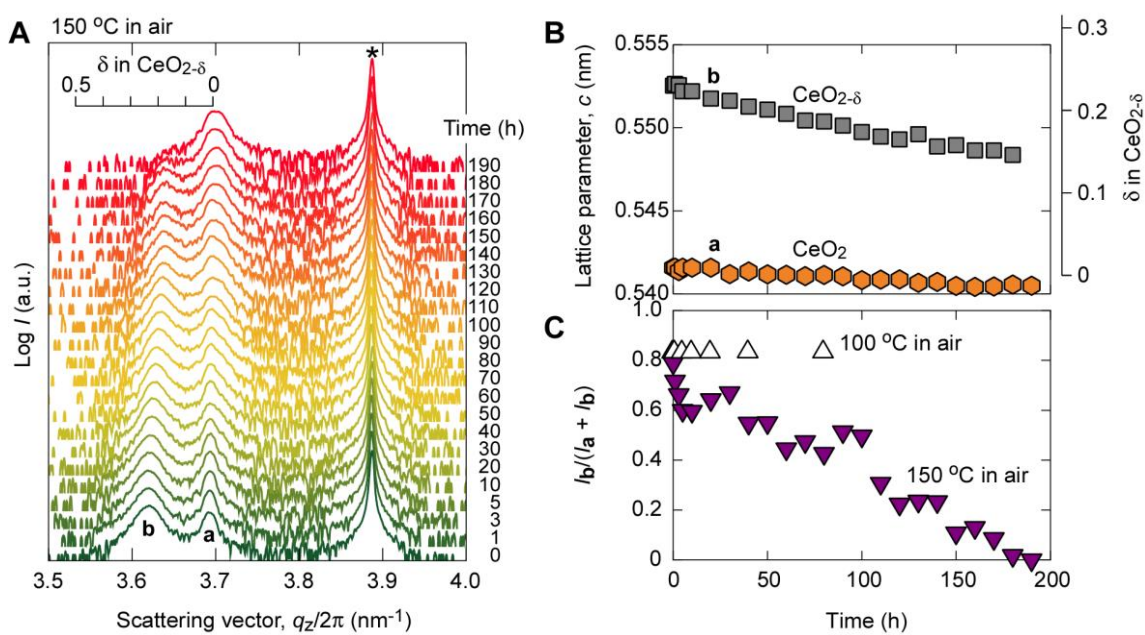

**Figure 3. Thermal stability of the reduced CeO$_2$-based thermal switch.** (**A**) Change in the out-of-plane XRD patterns of the reduced CeO$_2$-based thermal switch while kept at 150 °C in the air. After the reduction ($Q = -1.5 \times 10^{22}$ cm$^{-3}$, 0 h), the diffraction peak of 002 CeO$_{2-\delta}$ ($\delta$ ~0.24, **b**) is seen together with 002 CeO$_2$ (**a**). The peak position of **b** shifts to the higher $q_z/2\pi$ side and the peak **b** becomes weaker with time. On the other hand, peak **a** becomes stronger with time. After 190 h, the peak **b** disappears. (**B**) Change of lattice parameter, *c* of the CeO$_2$ (**a**) and CeO$_{2-\delta}$ (**b**) with time. The *c* of the **b** phase gradually decreases with time, while that of the **a** phase is almost constant. (**C**) XRD peak intensity ratio (*I***b**/(*I***a** + *I***b**)), which reflects the volume fraction of the **b** phase. The results at 100 °C in air are also shown. At 150 °C, the intensity ratio is ~0.8 at the beginning and decreases almost linearly with time, reaching zero after 190 h. In contrast, the intensity ratio is almost constant at 100 °C.

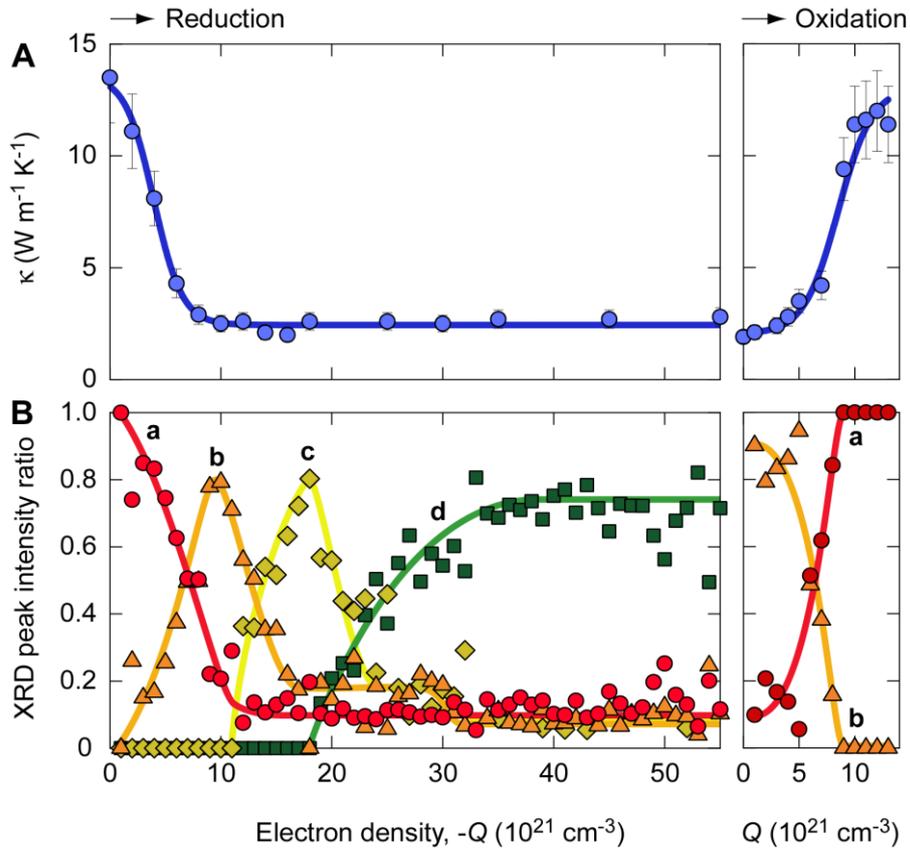

**Figure 4. Thermal conductivity modulation of the CeO₂-based thermal switches.**
(**A**) Thermal conductivity ($\kappa$) of the CeO$_2$ layer as a function of electron density ($Q$). The fully oxidized CeO$_2$ layer (**a**) shows a higher $\kappa$ of ~13 W m$^{-1}$ K$^{-1}$. The $\kappa$ decreases dramatically with $Q$ and reaches ~2.5 W m$^{-1}$ K$^{-1}$ at $Q$ ~10 × 10$^{21}$ cm$^{-3}$. Thereafter, $\kappa$ is nearly constant (~2.5 W m$^{-1}$ K$^{-1}$). On the other hand, when oxidized, $\kappa$ increases dramatically and reaches ~12.5 W m$^{-1}$ K$^{-1}$. (**B**) XRD peak intensity ratio. During reduction, the peak intensity ratio of **a** phase decreases dramatically with $Q$ and reaches ~0.1 when $Q$ ~10 × 10$^{21}$ cm$^{-3}$. On the other hand, the **b** phase increases and peaks around $Q$ ~10 × 10$^{21}$ cm$^{-3}$ and then decreases. Then, the **c** phase increases and has a peak around $Q$ ~18 × 10$^{21}$ cm$^{-3}$ and then decreases. Finally, the **d** phase increases and saturates around $Q$ ~35 × 10$^{21}$ cm$^{-3}$. During oxidation, the intensity of the **b** phase decreases dramatically and the intensity of the **a** phase increases dramatically. The thermal conductivity is nearly correlated with the volume fraction of the **a** phase.

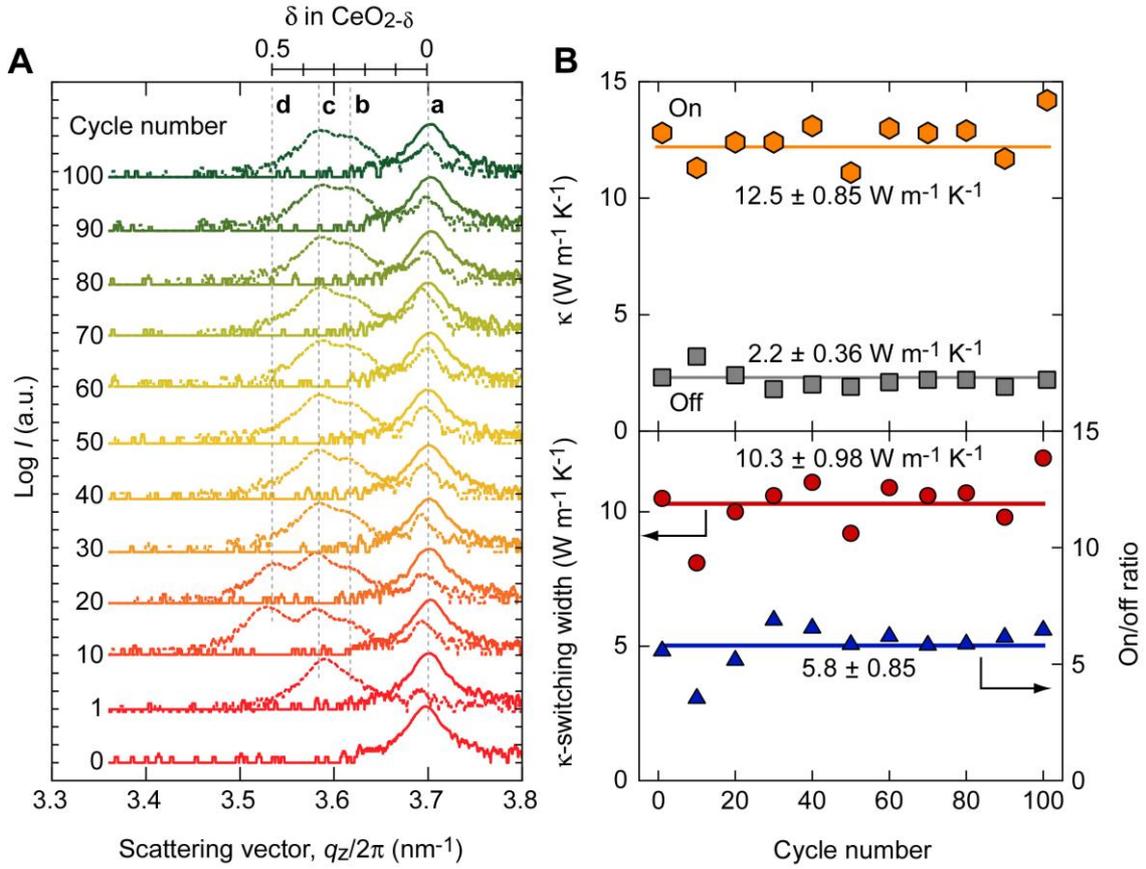

**Figure 5. Cyclic properties of the CeO$_2$-based thermal switch.** (**A**) Out-of-plane XRD patterns of the on-state (solid lines) and off-state (dashed lines) after each cycling. In the on-state, only the diffraction peak of the **a** phase (CeO$_2$) is observed, while in the off-state, diffraction peaks of reduced **b**−**d** phases with weak **a** phase are observed in all cases. (**B**) Changes in the thermal conductivity ($\kappa$) of the CeO$_2$ layer as a function of the number of redox cycles. The average $\kappa$ is 2.2 ±0.36 W m$^{-1}$ K$^{-1}$ for the reduced off-state and 12.5 ±0.85 W m$^{-1}$ K$^{-1}$ for the oxidized on-state. The average on/off $\kappa$-ratio of the CeO$_2$ layer is 5.8 ±0.85 and the $\kappa$-switching width is 10.3 ±0.98 W m$^{-1}$ K$^{-1}$.

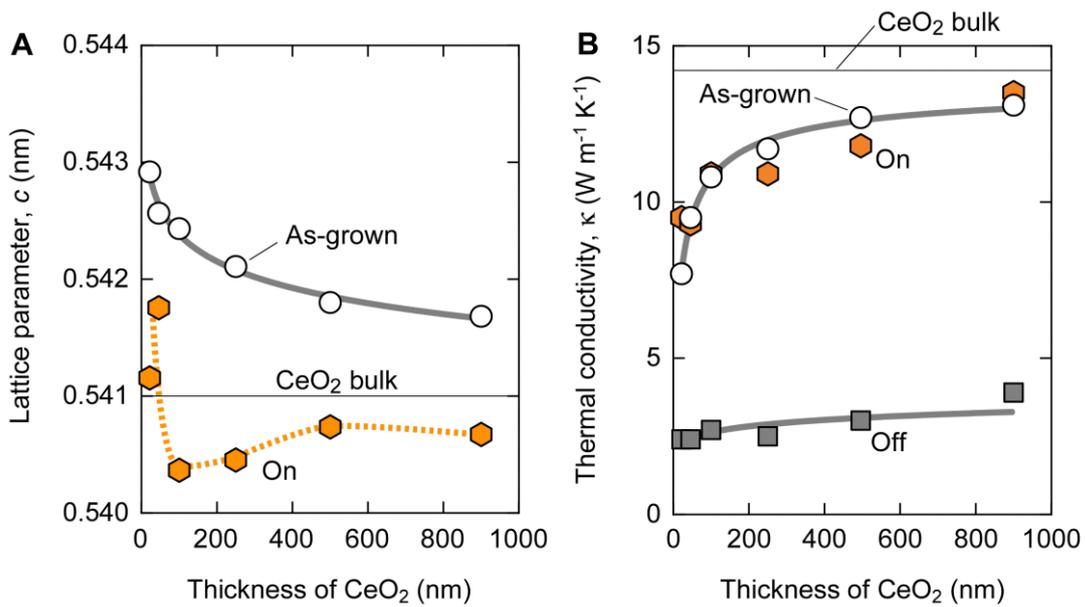

**Figure 6. Effect of the thickness of CeO$_2$ films.** (**A**) Change of lattice parameter, *c*, of as-grown and oxidized CeO$_2$ films with thickness. The lattice parameter decreases with thickness when the thickness is thin (~100 nm). (**B**) Variation of thermal conductivity of the CeO$_2$-based thermal switch measured at room temperature with CeO$_2$ thickness. The thermal conductivity increases dramatically with thickness in both the as-grown and on-state when the thickness is thin.

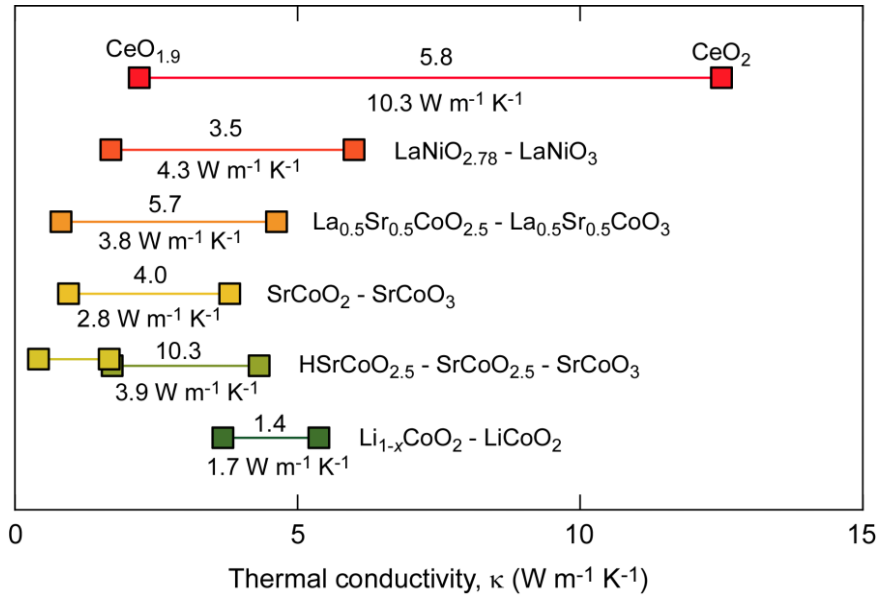

**Figure 7. Comparison of the thermal conductivity switching for the reported oxide-based electrochemical thermal switches.** $LiCoO_2/Li_{1-x}CoO_2$ (Cho *et al.* (*10*) 2014), $SrCoO_3/SrCoO_{2.5}/HSrCoO_{2.5}$ (Lu *et al.* (*12*) 2020), $SrCoO_3/SrCoO_2$ (Yang *et al.*(*15, 16, 28*) 2023), $La_{0.5}Sr_{0.5}CoO_3/La_{0.5}Sr_{0.5}CoO_{2.5}$ (Zhang *et al.*(*17*) 2023), $LaNiO_3/LaNiO_{2.78}$ (Bian *et al.*(*18*) 2024), $CeO_2/CeO_{1.9}$ (this study). The value above each line is the on/off *κ*-ratio and the value below each line is the *κ*-switching width. The $CeO_2$-based thermal switch shows a rather large on/off *κ*-ratio of 5.8 and very large *κ*-switching width of ~10.3 W m$^{-1}$ K$^{-1}$.

Supplementary Materials

# High-performance solid-state electrochemical thermal switches with earth-abundant cerium oxide

Ahrong Jeong[1,#], Mitsuki Yoshimura[2,#], Hyeonjun Kong[2], Zhiping Bian[2], Jason Tam[3], Bin Feng[3], Yuichi Ikuhara[3], Takashi Endo[1], Yasutaka Matsuo[1], and Hiromichi Ohta[1,*]

[1] *Research Institute for Electronic Science, Hokkaido University, N20W10, Kita, Sapporo 001-0020, Japan*
[2] *Graduate School of Information Science and Technology, Hokkaido University, N14W9, Kita, 060-0814 Sapporo, Japan*
[3] *Institute of Engineering Innovation, The University of Tokyo, 2-11-16 Yayoi, Bunkyo, Tokyo, 113-8656 Japan*

[#]Contributed equally

*Corresponding author
Hiromichi Ohta (hiromichi.ohta@es.hokudai.ac.jp)

**Keywords**

solid-state thermal switches, thermal conductivity, redox treatment, cerium oxide, electrochemistry, switching width

**Table of Contents**

S1. Fabrication of $CeO_2$-based solid-state electrochemical thermal switches
S2. Electrochemical reduction/oxidation of the $CeO_2$-based thermal switches
S3. Homologous phases of $Ce_nO_{2n-2}$ ($n \geq 4$, integer)
S4. STEM-EELS analysis of the $CeO_2$-based thermal switches
S5. TDTR simulation of the $CeO_2$-based thermal switches
S6. Thickness dependence of the $CeO_2$-based thermal switches

## S1. Fabrication of CeO$_2$-based solid-state electrochemical thermal switches

First, we fabricated CeO$_2$ epitaxial films on (001) oriented YSZ single crystal substrates by pulsed laser deposition (PLD) at 800 °C under an oxygen atmosphere (3 × 10$^{-3}$ Pa). Details of the CeO$_2$ film growth are described elsewhere. (*1, 2*) During the CeO$_2$ film growth in the PLD chamber, we monitored the reflection high-energy electron diffraction (RHEED, acceleration voltage: 25 kV, azimuth: <100>) patterns as shown in **Fig. S1A**. The rod-like RHEED pattern indicates that rather flat CeO$_2$ film is heteroepitaxially grown on the (001) YSZ substrate. The atomic force microscopy (AFM) topographic image of the film surface (**Fig. S1B**) confirms this conclusion. The film is composed of grains in the range of 30 – 100 nm. There are also many dark spots indicating threading dislocations. (*3*) **Figure S1C** shows the X-ray reflectivity of the CeO$_2$ film on (001) YSZ single crystal substrate. The calculated result shows that the root mean square roughness ($R_{rms}$) of the CeO$_2$ film is 0.72 nm and the thickness is 103 nm calculated from the Kiessig fringes.

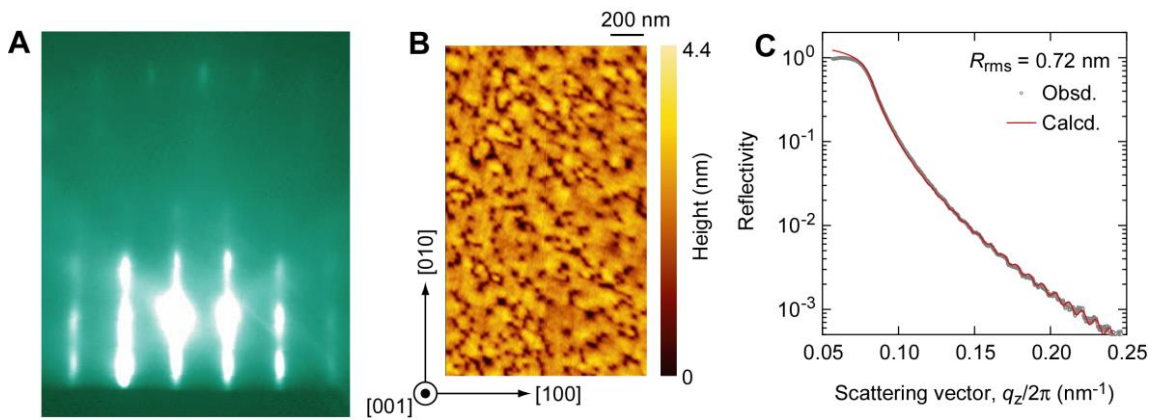

**Figure S1. Surface morphology of the CeO$_2$ film on (001) YSZ substrate.** (**A**) RHEED pattern (azimuth: <100>), (**B**) Topographic AFM image, (**C**) XRR pattern.

After checking the XRR, we deposited Pt films on the surface of the CeO$_2$ film and the back side of the YSZ substrate by DC magnetron sputtering, and then measured the out-of-plane XRD patterns of the film. **Figure S2A** shows the out-of-plane XRD pattern of the CeO$_2$-based thermal switch. Intense diffraction peaks of 002 and 004 CeO$_2$ are seen with 002 YSZ. As shown in the magnified pattern (**Fig. S2B**), Pendellösung fringes are observed, indicating a strong 00*l* orientation of the CeO$_2$ film. The film thickness that extracted from the Pendellösung fringes is 103 nm, which agrees well with the thickness extracted from the Kiessig fringes (**Fig. S1C**). The full width at half maximum (FWHM) of the X-ray rocking curve of 002 CeO$_2$ is 0.026 degrees, nearly equal to the resolution (**Fig. S2C**), confirming the strong 00*l* orientation of the film. We also measured the X-ray reciprocal space mapping (RSM) around 113 diffraction spots of $CeO_2$ and YSZ (**Fig. S2D**). The RSM indicates that a fully relaxed $CeO_2$ was grown on the (001) YSZ substrate.(*4*) The lateral grain size (*D*) was ~60 nm, which was estimated from the diffraction spot of 113 $CeO_2$ (**Fig. S2E**), in good agreement with the AFM observation (**Fig. S1B**).

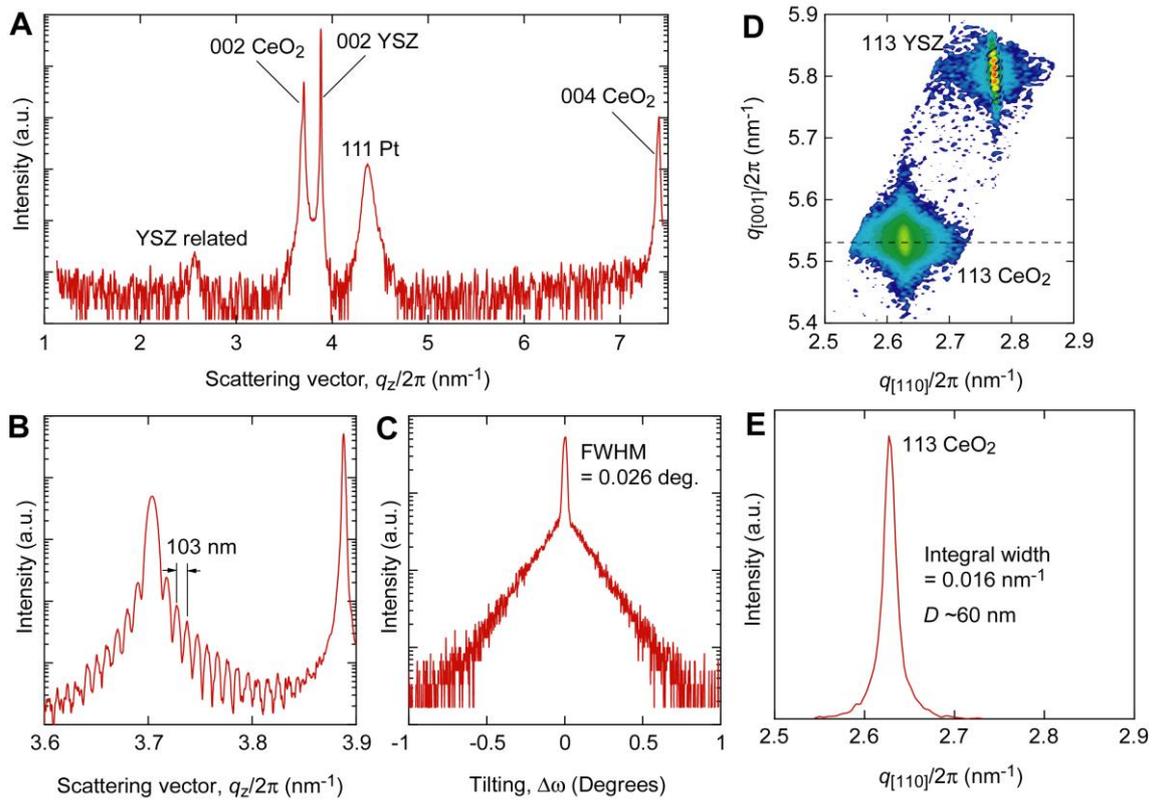

**Figure S2. X-ray diffraction analyses of the $CeO_2$-based thermal switch.** (**A**) Out-of-plane X-ray Bragg diffraction pattern, (**B**) Magnified XRD pattern of (**A**), (**C**) Out-of-plane X-ray rocking curve, (**D**) RSM, (**E**) Cross section pattern of (**D**) dotted line.

The layered structure of Pt/CeO$_2$/YSZ (**Fig. S3**) is visualized by the elemental mapping of the HAADF-STEM images (different sample).

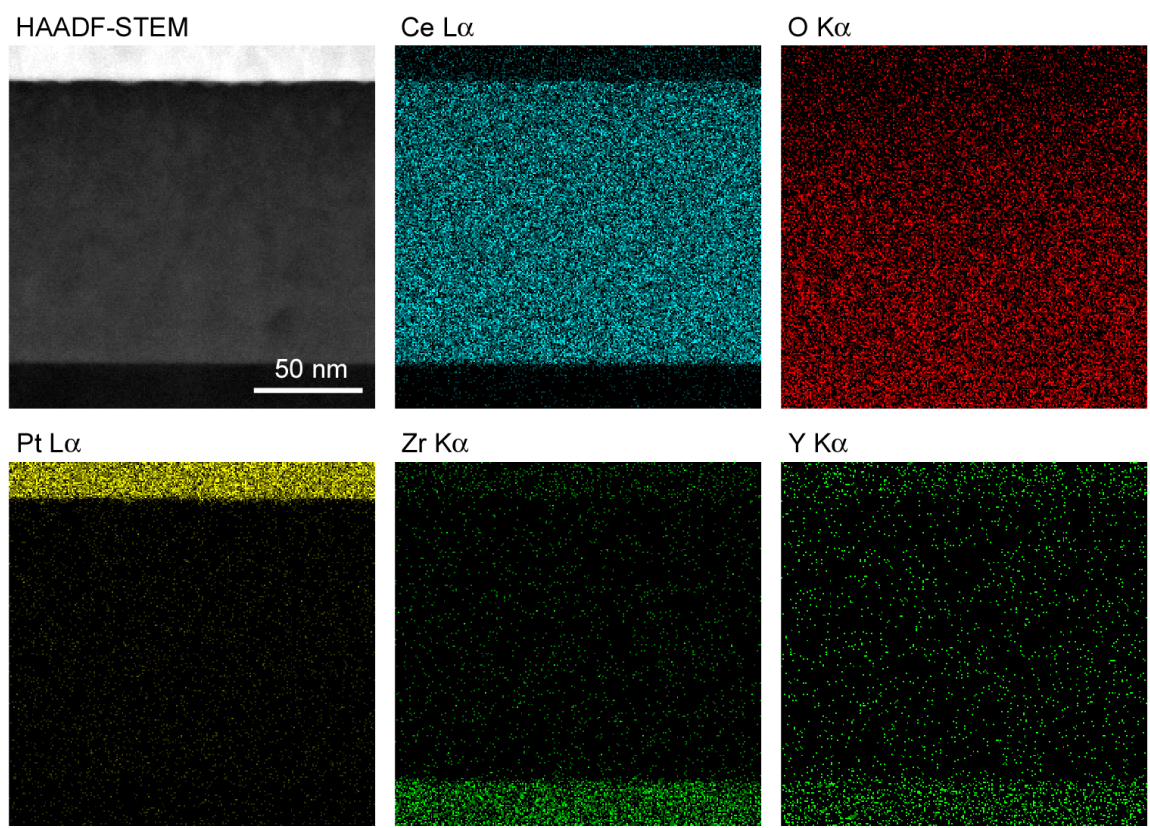

**Figure S3. EDS elemental mapping of the CeO$_2$-based thermal switch.** The layered structure composed of Pt/CeO$_2$/YSZ is visible.

## S2. Electrochemical reduction/oxidation of the CeO₂-based thermal switches

The electrochemical reduction/oxidation treatments were performed by applying a constant current of −10 μA/+10 μA (**Figs. 2A** and **2D**). These treatments were performed at an electron density $Q$ of $1 \times 10^{21}$ cm$^{-3}$ in each step. During electrochemical reduction (**Fig. 2A**), the absolute value of the applied voltage remains almost constant (~3.3 V). We repeated the reduction treatment until the total $Q$ reached $-5.5 \times 10^{22}$ cm$^{-3}$, corresponding to a total reduction time of 2255 s. In contrast, the voltage required for initial oxidation (**Fig. 2D**) was negative, reflecting spontaneous oxidation. When $Q$ exceeds ~$7 \times 10^{21}$ cm$^{-3}$, the required voltage increases dramatically from ~0 V to ~5 V. The required voltage became saturated when $Q = 1 \times 10^{22}$ cm$^{-3}$ was applied. After that, the required voltage decreased with time and became constant around 3.3 V.

In addition to the out-of-plane XRD measurements (**Figs. 2B** and **2E**), we measured X-ray reciprocal space mapping (RSM) of the CeO₂-based thermal switches after oxidation (**Fig. S4A**), after weak reduction (**Fig. S4B**), and after complete reduction (**Fig. S4C**). The RSM results indicate that all the Ce$_n$O$_{2n-2}$ ($n \geq 4$, integer, marked as **a**, **b**, **c**, and **d**) crystals are isotropic. Thus, all the Ce$_n$O$_{2n-2}$ crystals are fully relaxed and there is no epitaxial strain at the the film-substrate heterointerface. This was confirmed by the HAADF-STEM obsrevations (**Figs. S5A** and **S5B**). Misfit dislocations (yellow arrows) resulting from the lattice mismatch (~+5%) are visualized.

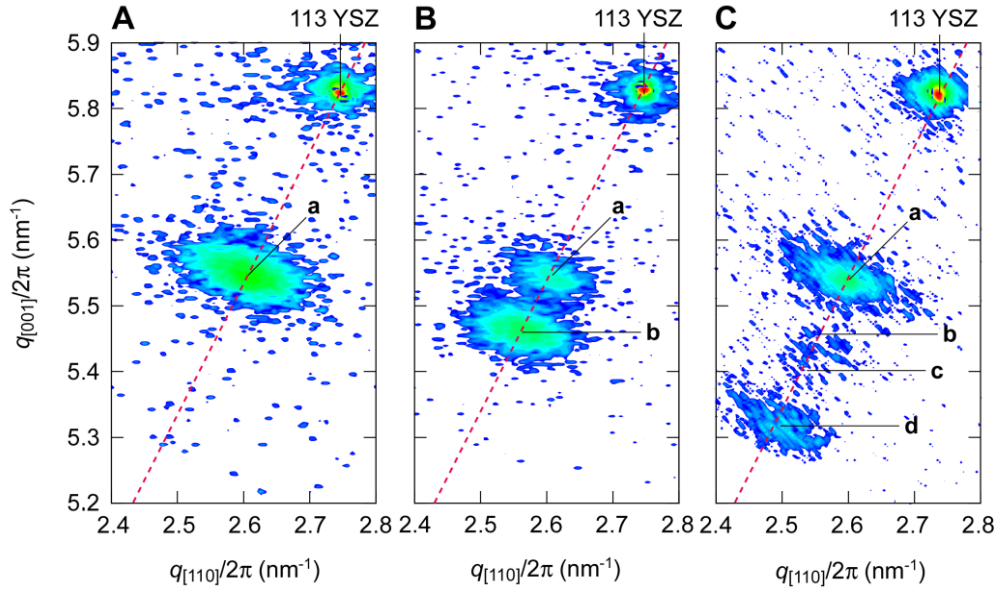

**Figure S4. X-ray reciprocal space mappings of the CeO$_2$-based thermal switch.** (**A**) After oxidation ($Q = 1.0 \times 10^{22}$ cm$^{-3}$), (**B**) after weak reduction ($Q = -1.4 \times 10^{22}$ cm$^{-3}$), (**C**) after complete reduction ($Q = -5.5 \times 10^{22}$ cm$^{-3}$). Red dotted lines with a slope of 2 are drawn across the diffraction peak of 113 YSZ; The crystal is isotropic when the diffraction peak is on the red dotted line. Diffraction peaks of **a**, **b**, **c**, and **d** phases are on the lines.

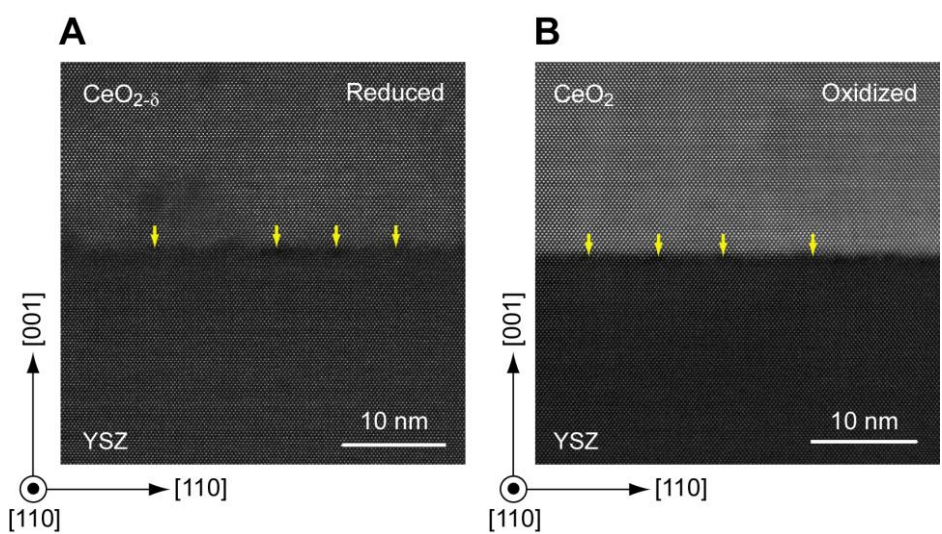

**Figure S5. HAADF-STEM images.** (**A**) after reduction (**b** phase), (**B**) after oxidation (**a** phase). Misfit dislocations (yellow arrows) resulting from the lattice mismatch (~+5%) are visualized.

### S3. Homologous phases of Ce$_n$O$_{2n-2}$ ($n \geq 4$, integer)

To analyze the homologous phases of **a**, **b**, **c**, and **d**, we plotted the lattice parameter (fluorite structure) of Ce$_n$O$_{2n-2}$ ($n \geq 4$, integer) as a function of $\delta$ in CeO$_{2-\delta}$. Using the reported lattice parameters (*5*) of CeO$_2$, CeO$_{1.832}$, and CeO$_{1.735}$, we found a linear relationship between the $\delta$ and the lattice parameter (**Fig. S6A**). The linear relationship was then extended to $\delta = 0.5$. Using the linear relationship, we clarified that the **a**, **b**, **c**, and **d** phases are CeO$_2$ ($n = \infty$), Ce$_9$O$_{16}$, Ce$_3$O$_5$, and Ce$_2$O$_3$, respectively. We also draw the crystal structures of CeO$_2$, Ce$_{11}$O$_{20}$, Ce$_7$O$_{12}$, and Ce$_2$O$_3$ using the crystallographic data reported by Kummerle *et al.*(*5*) The VESTA program (*6*) was used for the drawing and simulating the powder diffraction patterns (**Fig. S6B**). Except for CeO$_2$, the displacement of Ce and O atoms is large. This would reflect the low thermal conductivity of Ce$_n$O$_{2n-2}$ ($n \geq 4$, integer) except CeO$_2$.

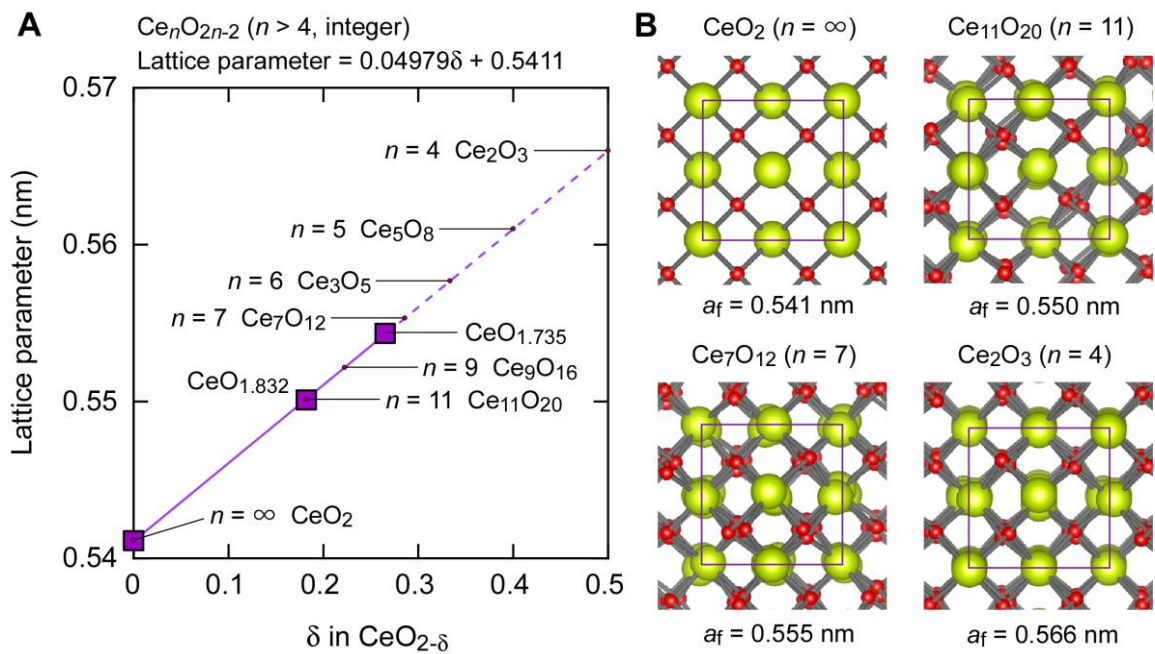

**Figure S6. Homologous phases of $Ce_nO_{2n-2}$.** (**A**) Lattice parameters (fluorite structure) of $Ce_nO_{2n-2}$ ($n \geq 4$, integer) as a function of $\delta$ in $CeO_{2-\delta}$. First, the lattice parameters of the reported $CeO_2$, $CeO_{1.832}$, and $CeO_{1.735}$ were plotted. Then, the solid line was drawn with the least square fit. The line then was extended to $\delta = 0.5$ by assuming a linear relationship. Multiple $n$-values are indicated on the line. (**B**) Crystal structures of $CeO_2$, $Ce_{11}O_{20}$, $Ce_7O_{12}$, and $Ce_2O_3$, that drawn using the crystallographic data reported by Kummerle *et al.*(*5*) The VESTA program (*6*) was used to draw and simulate the powder diffraction patterns. (Yellow balls: Ce, Red balls: Oxygen)

### S4. STEM-EELS analysis of the CeO₂-based thermal switches

To clarify that the reduction of $Ce^{4+}$ to $Ce^{3+}$ occurs after the electrochemical reduction treatment, we performed the STEM-EELS analyses (**Fig. S7**) of the weakly reduced sample ($Q = -1.0 \times 10^{22}$ cm$^{-3}$). It is known that the intensity ratio of $M_5/M_4$ of Ce $M_{4,5}$ edge EELS spectra reflects the $Ce^{3+}/Ce^{4+}$ concentration in $CeO_{2-\delta}$ films ($Ce^{4+}$: $M_5/M_4 = 0.90$, $Ce^{3+}$: $M_5/M_4 = 1.25$)(*1*). Ideally, the equation of $M_5/M_4$ ratio $= 0.9 + 0.7\delta$ ($0 \leq \delta \leq 0.5$) is established. The $M_5/M_4$ ratio was calculated from the positive part of the second derivative of the experimental spectra. The $M_5/M_4$ ratio of the reduced $CeO_{2-\delta}$ film was 0.99, while that of the oxidized $CeO_2$ was 0.90. Thus, $\delta$ in the $CeO_2$ film was estimated to be 0.13 (**b** phase). Note that the observed $M_5/M_4$ ratio was varied as the TEM sample thickness, since the reduced film was composed of **a** and **b** phases.

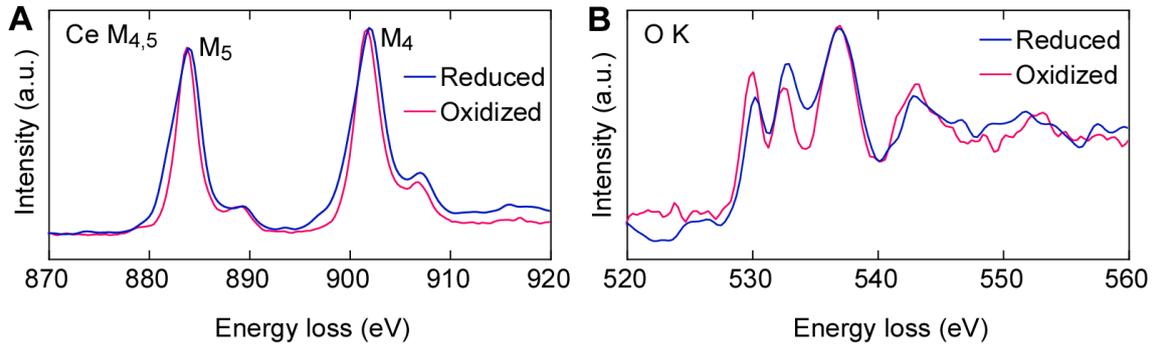

**Figure S7. EELS of the CeO₂-based thermal transistor.** (**A**) EELS spectra of Ce $M_{4,5}$ edges. (**B**) EELS spectra of O K edges. The $M_5/M_4$ ratio of the reduced $CeO_{2-\delta}$ film was 0.99, whereas that of the oxidized $CeO_2$ was 0.90.

## S5. TDTR simulation of the CeO₂-based thermal switches

The cross-plane thermal conductivity of the CeO₂-based thermal switches was measured by the TDTR method at room temperature. We used commercially available TDTR equipment (PicoTR, PicoTherm Co.(*7*)). Sputtered Pt films were used as transducers. The samples were irradiated with femtosecond laser pulses (wavelength: 1550 nm; pulse duration: 0.5 ps; laser spot size: 45 μm), and the change in the reflectivity in the time domain was recorded with a probe laser (wavelength: 775 nm; pulse duration: 0.5 ps; laser spot size: 25 μm). The obtained thermoreflectance phase signals (**Figs. S8 – S10**) were analyzed with a software package provided by the manufacturer (PicoTherm Co.(*7*)) using the physical properties listed in **Table S1**. We assumed that the specific heat capacity ($C_p$) of CeO₂ is always 358 J kg⁻¹ K⁻¹ and it is independent on the oxygen deficiency. To avoid the overestimation of $\kappa$, the Kapitza resistance ($R_K$) values at the Pt/CeO₂ interface lower than $4.0 \times 10^{-9}$ m² K W⁻¹ were used.

**Table S1. Simulation condition of the TDTR decay curves for the CeO₂-based thermal switches.** $C_p$: specific heat capacity, $\kappa$: thermal conductivity, $R_K$: interfacial thermal resistance.

|  | Pt | CeO₂ | YSZ |
| --- | --- | --- | --- |
| Density (kg/m³) | 21500 | 7215 | 5770 |
| $C_p$ (J/kgK) | 133 | 358 | 460 |
| $\kappa$ (W/mK) | 46 | ---- | 2.00 |
| $R_K$ (m²K/W) | $1.0 \times 10^{-9} - 4.0 \times 10^{-9}$ | $1.0 \times 10^{-9}$ | |

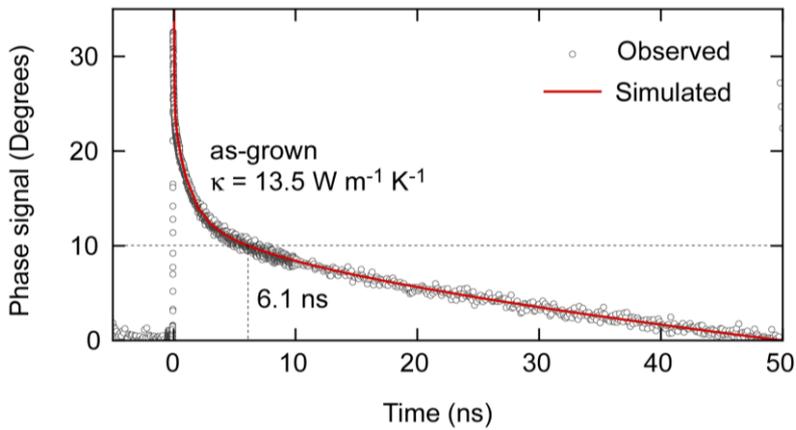

**Figure S8. Typical TDTR decay curve of the as-grown CeO₂-based thermal transistor.** The $R_K$ at the Pt/CeO$_2$ interface was $4.0 \times 10^{-9}$ m$^2$ K W$^{-1}$. The simulated curve reproduces the observed decay curve. The extracted $\kappa$ of the CeO$_2$ layer is 13.5 W m$^{-1}$ K$^{-1}$. The delay time at 10° is ~6.1 ns.

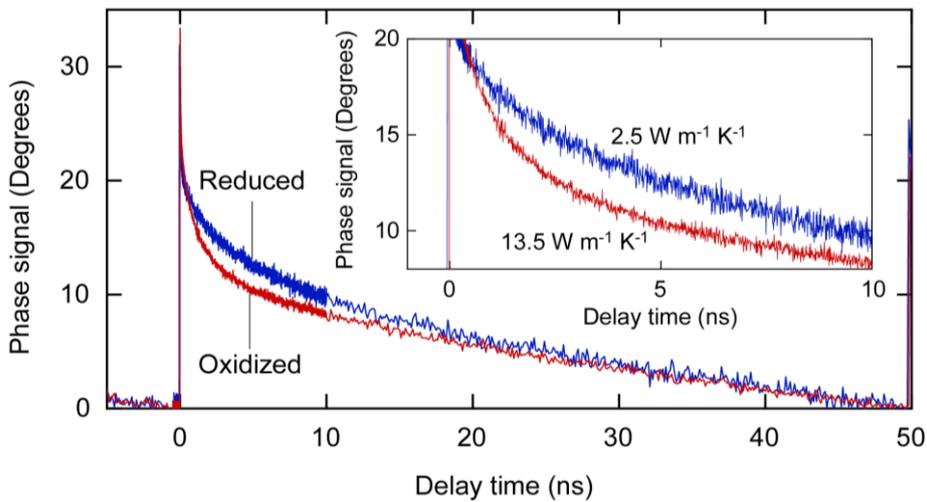

**Figure S9. Thermal conductivity modulation of the CeO₂-based thermal switch after oxidation and reduction.** TDTR decay curves of the CeO$_2$-based thermal switch measured at room temperature. The inset shows the magnified graph. The decay of the TDTR phase signal of the oxidized sample is faster than that of the reduced one.

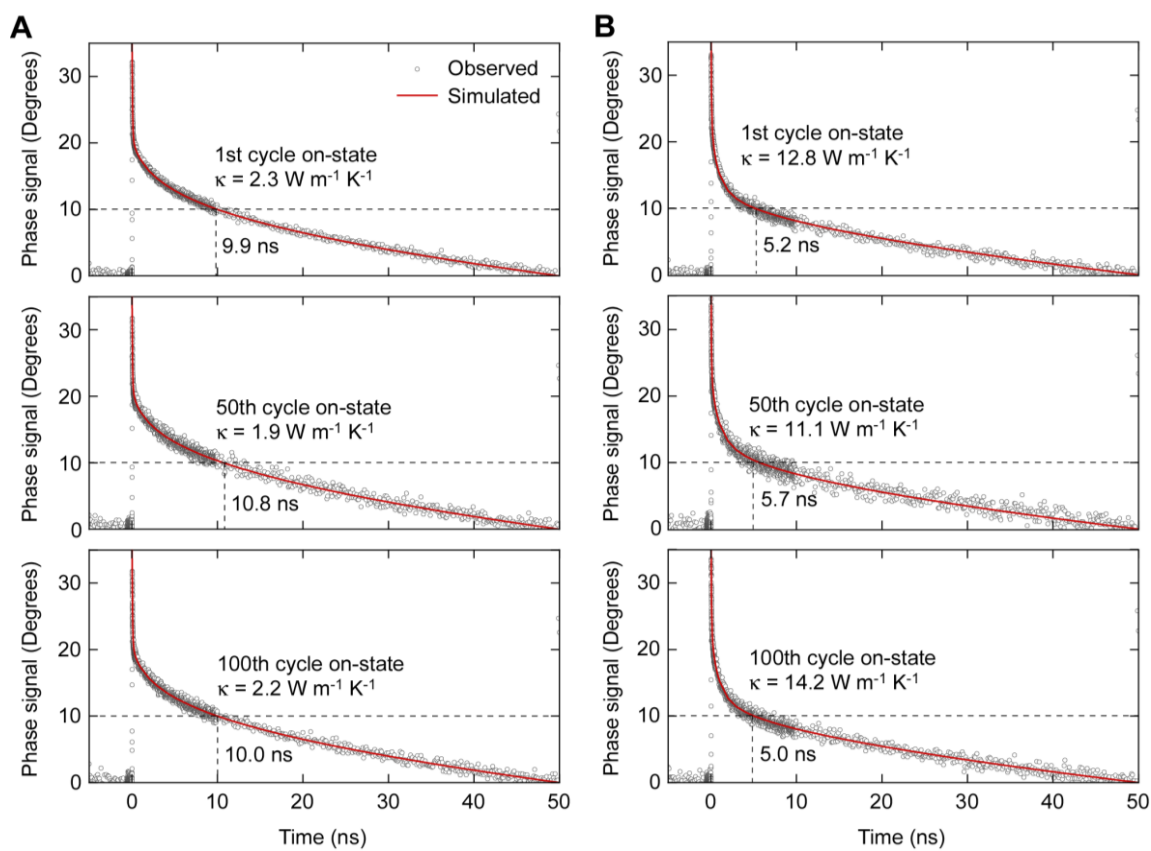

**Figure S10. TDTR decay curves of the CeO$_2$-based thermal switch after 1st, 50th, and 100th cycles.** (**A**) Reduced off-state, (**B**) Oxidized on-state. The delay time is shown for comparison.

### S6. Thickness dependence of the CeO$_2$-based thermal switches

To clarify the effect of CeO$_2$ thickness on the thermal switch performance, we fabricated several thermal switches with different CeO$_2$ film thicknesses. Electrochemical reduction of the thermal switches was performed by applying a constant DC current of −10 μA until the total $Q$ reached $-1.5 \times 10^{22}$ cm$^{-3}$. On the other hand, the oxidation treatment was performed by applying a constant DC current of +10 μA until the total $Q$ reached $0.9 \times 10^{22}$ cm$^{-3}$. **Figure S11** summarizes the out-of-plane XRD patterns of the resulting thermal switches. The diffraction peak of the **a** phase (CeO$_2$) of the as-grown samples shifts to a higher $q_z/2\pi$ side with increasing thickness. After the reduction treatment, **a**, **b**, and **c** phases are randomly distributed most likely due to the contribution of spontaneous oxidation. After the oxidation treatment, only **a** phase diffraction peaks are observed and the peaks shift slightly to a higher $q_z/2\pi$ side with increasing thickness.

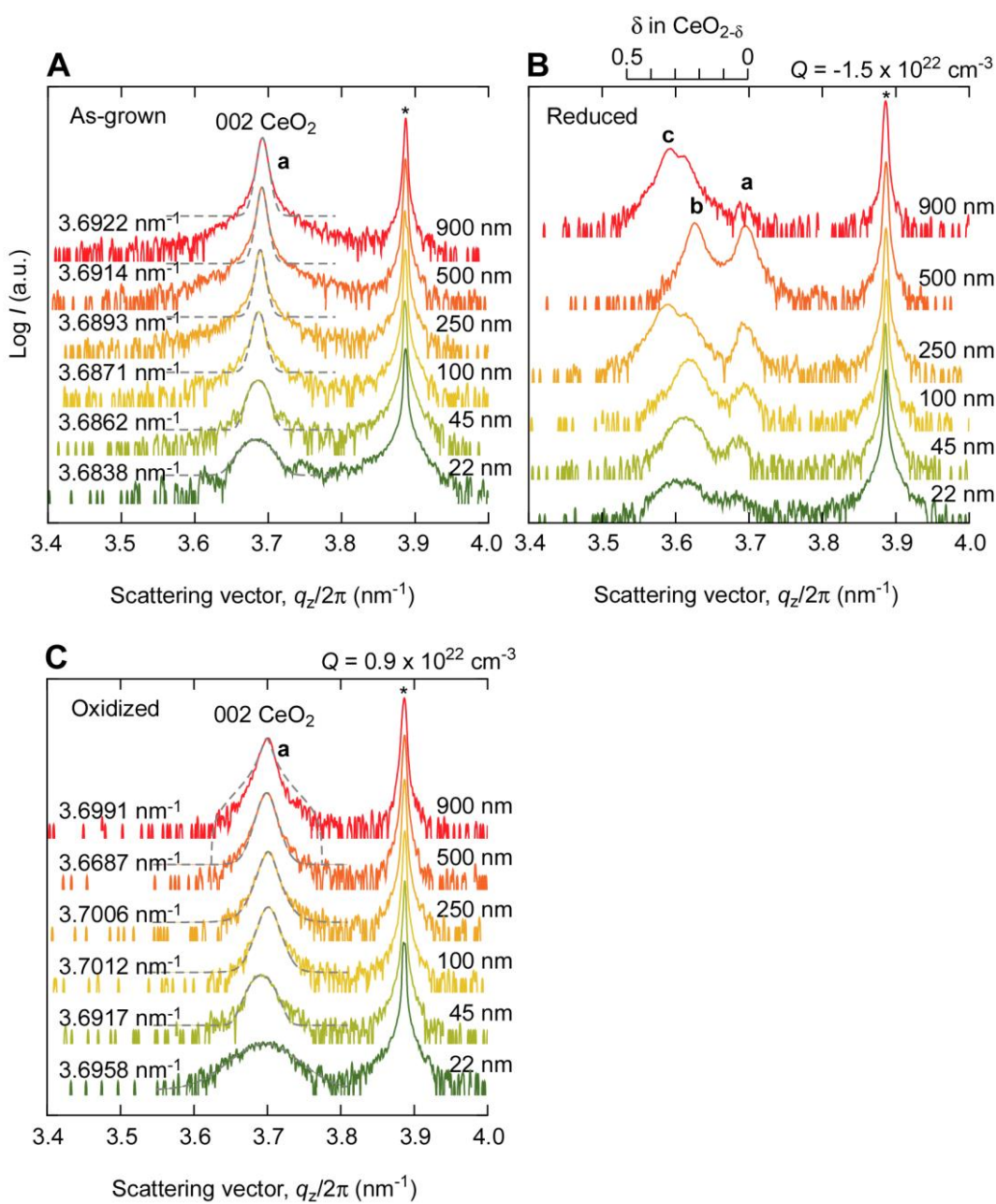

**Figure S11. Out-of-plane XRD patterns of CeO$_2$-based thermal switches with varied CeO$_2$ thickness.** (**A**) As-grown, (**B**) reduced ($Q = -1.5 \times 10^{22}$ cm$^{-3}$), (**C**) oxidized ($Q = 0.9 \times 10^{22}$ cm$^{-3}$). (**A**, **C**) The diffraction peak of **a**-phase (CeO$_2$) shifts to a higher $q_z/2\pi$ side with increasing thickness. (**B**) Although $Q$ for the reduction is the same as $-1.5 \times 10^{22}$ cm$^{-3}$, **a**, **b**, and **c** phases are randomly distributed.